\documentclass{elsarticle}

\usepackage{lineno,hyperref}
\modulolinenumbers[100]

\journal{International Journal of Heat and Fluid Flow}

\usepackage{numcompress}\biboptions{authoryear}

\usepackage{lmodern}
\usepackage{geometry}
\usepackage{epsfig}
\usepackage{epsf}
\usepackage{psfrag}
\usepackage{pstool}
\usepackage{tikz}
\usepackage{textcomp}
\usepackage{trimclip}
\usepackage{amssymb}
\usepackage{trimclip} 
\usepackage{siunitx} 
\usepackage{graphicx}
\usepackage{epstopdf, epsfig}
\usepackage[utf8]{inputenc}
\usepackage{upgreek}
\usepackage{booktabs}
\usepackage{graphics} 
\usepackage{xcolor}
\usepackage{amsmath}
\usepackage{mwe} 
\usepackage{enumitem}

\usepackage{multirow}
\usepackage{mathrsfs,amsmath}
\usepackage{bm}
\usepackage{multirow,bigdelim}
\usepackage{array,multirow}
\usetikzlibrary{arrows.meta}
\usepackage{color}
\usepackage{threeparttable}	
\usepackage{dcolumn}		
\newcolumntype{d}[1]{D{.}{.}{#1}}
\usetikzlibrary{decorations}
\usetikzlibrary{decorations.markings}
\usetikzlibrary{calc,decorations.pathreplacing}
\usepackage{mathtools}
\usepackage{xfrac}
\usepackage{ar}
\usepackage{enumitem}
\usepackage{url}
\usepackage{wasysym}

\def\AR{\clipbox{0pt 0pt .32em 0pt}\AE\kern-.30emR}

\newcommand*{\myfont}{\fontfamily{ptmri}\selectfont}
\DeclareTextFontCommand{\textmyfont}{\myfont}

\begin{document}
\sisetup{group-separator={,}}
\sloppy

\begin{frontmatter}

\title{Near wall coherence in wall-bounded flows and implications for flow control}

\author[mymainaddress,mysecondaddress]{M. Samie\corref{mycorrespondingauthor}}
\cortext[mycorrespondingauthor]{Corresponding author}
\ead{m.samie@queensu.ca}

\author[mythirdaddress]{W. J. Baars}
\author[mysecondaddress]{A. Rouhi}
\author[myfourthaddress]{P. Schlatter}
\author[myfourthaddress]{R. \"Orl\"u}
\author[mysecondaddress]{I. Marusic}
\author[mysecondaddress]{N. Hutchins}
\address[mymainaddress]{Department of Mechanical and Materials
   Engineering,
    Queen's University,
    Kingston, ON, Canada K7L 3N6}
\address[mysecondaddress]{Department of Mechanical Engineering, University of Melbourne, Victoria 3010, Australia}
\address[mythirdaddress]{Faculty of Aerospace Engineering, Delft University of Technology, 2629 HS Delft, The Netherlands}
\address[myfourthaddress]{ Linn\'e FLOW Centre, KTH Mechanics, Royal Institute of Technology, Stockholm, Sweden}

\begin{abstract}
Opposition-control of the energetic cycle of near wall streaks in wall-bounded turbulence, using numerical approaches, has shown promise for drag reduction. For practical implementation, opposition control is only realizable if there is a degree of coherence between the sensor--actuator pairs of the control system (and for practicality these sensors and actuators should typically be wall-based to avoid parasitic drag). As such, we here inspect the feasibility of real-time control of the near-wall cycle, by considering the coherence between a measurable wall-quantity, being the wall-shear stress fluctuations, and the streamwise and wall-normal velocity fluctuations in a turbulent boundary layer. Synchronized spatial and temporal velocity data from two direct numerical simulations and a fine large eddy simulation at $Re_\tau \approx 590$ and $ 2000$ are employed. This study shows that the spectral energy of the streamwise velocity fluctuations that is stochastically incoherent with wall signals is independent of Reynolds number in the near wall region (up to the viscous-scaled wall-normal height $z^+ \approx 20$). Consequently, the streamwise energy-fraction that is stochastically wall-coherent grows with Reynolds number due to the increasing range of energetic large scales. This thus implies that a wall-based control system has the ability to manipulate a larger portion of the total turbulence energy at off-wall locations, at higher Reynolds numbers, while the efficacy of predicting/targeting the small scales of the near-wall cycle remains indifferent with varying Reynolds number. Coherence values of 0.55 and 0.4 were found between the streamwise and wall-normal velocity fluctuations at the near wall peak in the energy spectrogram, respectively, and the streamwise fluctuating friction velocity. These coherence values, which are considerably lower than 1 (maximum possible coherence) suggest that a closed-loop drag reduction scheme targeting near wall cycle of streaks alone (based on sensed friction velocity fluctuations) will be of limited success in practice as the Reynolds number grows.
\end{abstract}


\end{frontmatter}

\linenumbers

\section{Introduction}
Given that skin-friction drag constitutes a large fraction of the total aerodynamic drag of transport systems (e.g. ships, aircraft and piping systems), a small percentage of skin-friction drag reduction is greatly rewarded both economically and environmentally. It has been known for decades that turbulent flows are comprised of several types of coherent structures \citep[e.g.][]{kline1967structure, brown1977large, townsend1980structure}. Quasi-streamwise vortices (QSVs) and streaks--primarily residing in the buffer layer--are recognized to be strongly associated with high skin-friction and hence turbulent wall drag \citep{orlandi1994generation}. Therefore, the majority of research aiming at skin-friction drag reduction has focused on suppressing the QSVs and streaks \citep[e.g.][]{moin1994feedback, choi1994active, lee1998suboptimal, choi1998turbulent, rathnasingham2003active, bai2014active, qiao2017turbulent, toedtli2019predicting}. For engineering systems, these near-wall features are physically small; e.g. the average diameter of the QSVs is $\sim 0.1$ mm near the fuselage of a passenger airplane in cruise. Recent advances in microelectromechanical systems technologies have provided sufficiently small sensors and actuators to target such small structures \citep{kasagi2009microelectromechanical}, but a continuous lay-up of micro-sensors/actuators is practically infeasible when large-scale transport systems are concerned. Attempts have also been made to reduce drag by manipulating large-scale coherent structures in the logarithmic and outer regions of turbulent boundary layers (TBLs) \citep[e.g.][]{schoppa1998large,abbassi2017skin}. In a direct numerical simulation (DNS) of turbulent channel flow at $Re_\tau=180$, \cite{schoppa1998large} reported a drag reduction of up to 50\% by using spanwise jets as a large-scale flow forcing. However, there is ongoing debate over the effectiveness of large-scale friction control with increasing Reynolds number \citep{canton2016reynolds, deng2016origin, yao2017large, yao2018drag}. Additionally at $Re_\tau = 180$ there is no spectral scale separation between inner and outer scales. Here the friction Reynolds number is defined as $Re_\tau \equiv U_\tau \delta/\nu$, where $U_\tau \equiv \sqrt{\tau_0/\rho}$ is the mean wall-friction velocity ($\tau_0$ is the wall shear stress), $\delta$ is the boundary layer thickness and $\nu$ and $\rho$ are the fluid kinematic viscosity and density, respectively.

Strategies for turbulence skin-friction drag reduction are generally classified in two categories: passive control and active control \citep{gad1996modern}. In passive control a flow is modified without external energy inputs while in active control, a steady or unsteady modification is continuously applied to the flow. Active control is further divided into (predetermined) open-loop and (reactive) closed-loop schemes. Under both schemes, the actuation may be adjusted based on sensor inputs that record the flow characteristics \citep{brunton2015closed}. For a practical control system that aims at reducing skin-friction, sensors and actuators need to be wall-based and flush-mounted to avoid parasitic drag. Consequently, there is an unavoidable wall-normal separation, $\Delta z$ (coordinate $z$ denotes the wall-normal direction), between the sensor and flow structures that are targeted, in say the buffer layer  \citep[e.g.][]{choi1994active, qiao2018turbulent} or the outer layer \citep[e.g.][]{abbassi2017skin}. Moreover, a separation in the streamwise direction, $\Delta x$, must often be applied between the sensor and actuator due to the infeasibility of colocation and also to take into account the controller processing time and the time delay associated with the actuator response \citep[see for example the experimental set-up in][]{rebbeck2006wind}. The $\Delta z$ and $\Delta x$ separations, in combination with the evolution of turbulence, results in a loss of correlation between the measured signal and the structures being targeted. The degree of coherence between the signals is critical for determining the feasibility of implementing meaningful opposition control hardware. Linear coherence spectra (explained in detail in \S \ref{sec:feasibility}), inspecting a scale-by-scale normalized correlation between the two signals, proves useful for this purpose. 
\begin{figure}
\includegraphics[width=1.00\textwidth]{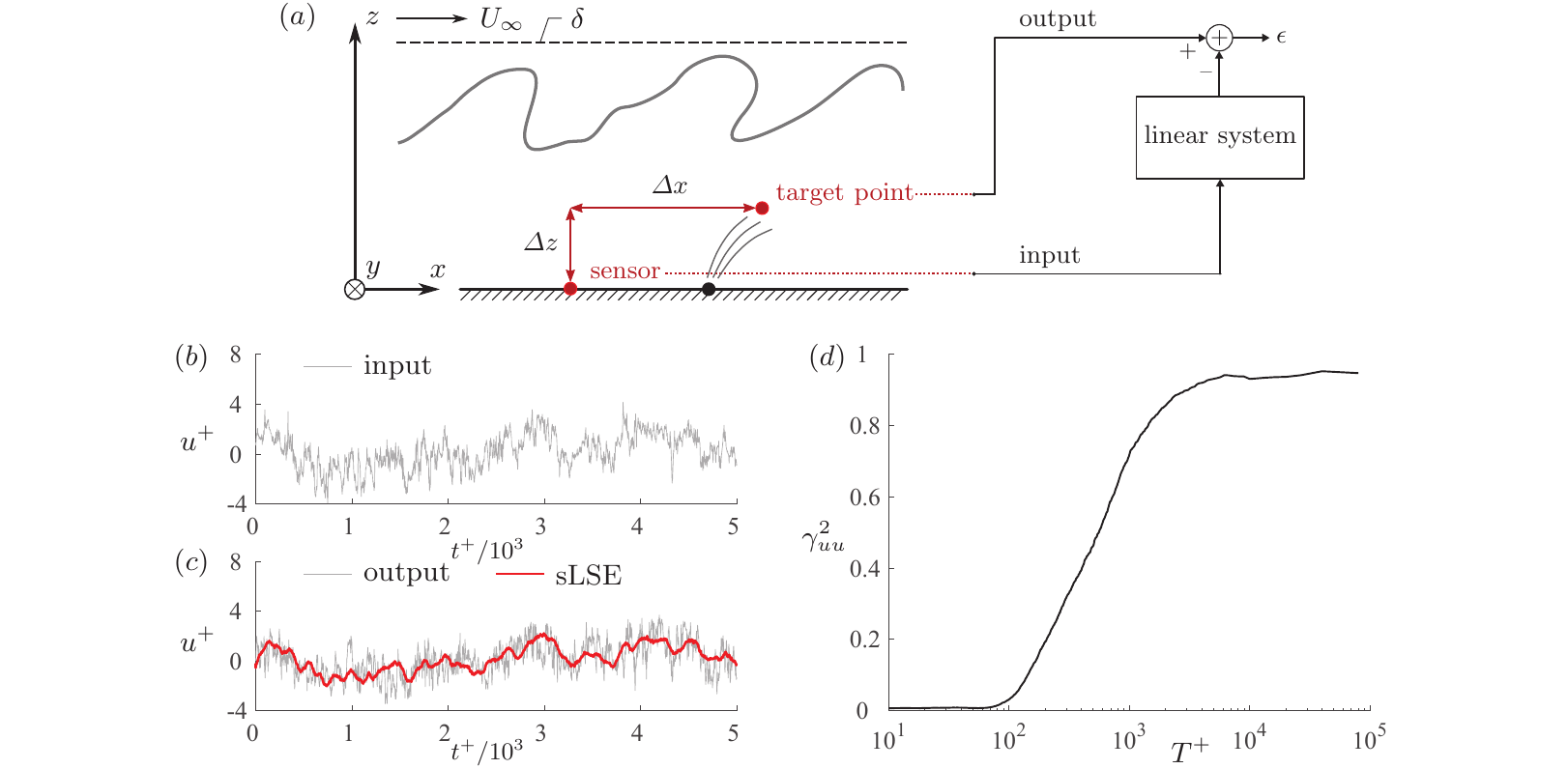}
\caption{($a$) Illustration of a wall-based sensor and target point for real-time flow control in a turbulent boundary layer. ($b,c$) Input and output time series of $u$-fluctuations at two separated points. The spectral Linear Stochastic Estimate (sLSE) of the output, generated from the input signal, is shown in red. ($d$) Linear coherence spectra corresponding to the input--output time series (the spectral scale is in terms of period $T^+ \equiv 1/f^+$).}
   \label{fig:ischem}
\end{figure}

\subsection{Feasibility of opposition control} \label{sec:feasibility}
This paper addresses the limitations of near wall opposition control in wall-bounded turbulence. With a wall-based sensor (measuring an input signal for a controller) and an off-wall target point (e.g. where opposition control is anticipated), the goodness of the control system relies on how well the input signal represents the state of the target point. Figure~\ref{fig:ischem}(a) illustrates this scenario for an example control set-up \citep[e.g.][]{abbassi2017skin}. Generally, a routine action of the controller would be to predict the system output from the input, so that an actuator could properly act upon the target location. By linking the input and output via a linear time invariant system, the output estimate $u_{\rm sLSE}$ differs from the measured output $u_{\rm o}$ by an error $\epsilon$. The estimate itself can be generated via a spectral linear stochastic estimate \citep{bendat:1980bk,tinney:2006a}, following
\begin{equation}\label{eq:slse}
 \tilde{u}_{\rm sLSE}\left(f\right) = H_L\left(f\right) \tilde{u}_{\rm i}\left(f\right).
\end{equation}
Here, $H_L$ is a complex-valued linear transfer kernel and $\tilde{u}_{\rm sLSE}$ and $\tilde{u}_{\rm i}$ are the Fourier transforms of the estimated and input time series, $u_{\rm sLSE}(t)$ and $u_{\rm i}(t)$, respectively. The kernel $H_L$ is found from a calibration experiment with synchronized two-point data, via
\begin{equation}\label{eq:hl}
 H_L\left(f\right) = \frac{\langle \tilde{u}_{\rm o}\left(f\right) \overline{\tilde{u}_{\rm i}}\left(f\right) \rangle}{\langle \tilde{u}_{\rm i}\left(f\right) \overline{\tilde{u}_{\rm i}}\left(f\right) \rangle} = \vert H_L\left(f\right)\vert e^{j\phi\left(f\right)},
\end{equation}
where $\tilde{u}_o$ is the Fourier transform of the  measured output signal. The complex-valued kernel equals the input-output cross-spectrum, divided by the input spectrum, and thus includes the system gain and phase. Angled brackets ($\langle \rangle$) indicate ensemble averaging and an over line indicates the complex conjugate. Once a stochastic kernel is computed from synchronized two-point data, a time-domain estimate of the system is found via (\ref{eq:slse}) and the inverse Fourier transform:
\begin{equation}
 \label{eq:lse}
  u_{\rm sLSE}\left(t\right) = \mathcal{F}^{-1} \left[\tilde{u}_{\rm sLSE}(f)\right].
\end{equation}
Sample input and output time series from experimental data \citep{baars:2016ab} are shown in figures~\ref{fig:ischem}(b,c). From a comparison between the estimated time series, via (\ref{eq:lse}), and the measured time series, it is evident that the smaller scales are not accounted for. Specifically, for a certain input--output separation, there is a limited linear mechanism of energy coupling at the smaller scales \citep[e.g.][]{adrian:1987a,guezennec:1989a,naguib:2001a}, which manifests as an inability to estimate these scales. The linear coherence spectrum (LCS) proves useful in quantifying this energy coupling in a scale-by-scale manner and in a stochastic sense. For $u$ input and output signals, separated by $\Delta x$, $\Delta y$ and $\Delta z$ in the $x$, $y$, and $z$ directions, respectively, the LCS is defined as
\begin{equation}
\gamma_{uu_\tau}^2(\Delta x,\Delta y, \Delta z; f) \equiv \frac{\vert \langle \tilde{u}_{\rm o}(f) \overline{\tilde{u}_{\rm i}}(f)\rangle \vert ^2}{\langle \vert \tilde{u}_{\rm o}(f) \vert ^2 \rangle  \langle \vert \tilde{u}_{\rm i}(f) \vert ^2 \rangle} = \frac{\vert \phi_{u_{\rm oi}}(f) \vert ^2}{\phi_{u_{\rm ii}}(f) \phi_{u_{\rm oo}}(f)}.
\label{eq:lcs}
\end{equation}
Here, $\tilde{u}$ is again the Fourier transform of $u$ in either $x$ (for spatial data) or time (for temporal data) and $\phi_{u_{\rm oi}}(f)$, $\phi_{u_{\rm ii}}(f)$ and $\phi_{u_{\rm oo}}(f)$ denote the cross-spectrum and the input and output power spectra, respectively. Throughout this work, $\gamma_{uu_\tau}^2$ will be presented as a function of the streamwise wavelength $\lambda_x$. For temporal data that means that $\gamma_{uu_\tau}^2$ is calculated in the frequency domain and converted to the wavelength domain by invoking Taylor's frozen turbulence hypothesis using the local mean velocity at each $z$ as the convection velocity. The ratio $\gamma_{uu_\tau}^2$ is a per-scale, normalized correlation between two signals and is bounded by 0 (no coherence) and 1 (perfect coherence). Figure~\ref{fig:ischem}(d) presents the LCS for the input and output time series shown in figures~\ref{fig:ischem}(b,c). Indeed the LCS indicates the absence of coherence at the small scales ($T^+<100$) and a near-perfect coherence at the largest scales ($T^+>7000$). In a stochastic sense, the value of $\gamma^2_{uu_\tau}$ may be interpreted as the \emph{fraction of energy in the output signal} that can be estimated via an sLSE procedure from the input signal, since $\gamma^2_{uu_\tau} = \vert H_L(f) \vert^2 \phi_{u_{\rm ii}}/\phi_{u_{\rm oo}}$ (the estimated output energy--from the input--divided by the measured output energy).

Opposition control at the target location can, in the best  hypothetical scenario possible, only act upon the estimated signal. An estimate of an off-wall velocity signal, from wall-based quantities, can be achieved via different techniques. For instance, efforts can employ neural networks \citep{guemes2019sensing, guastoni2020use} or other correlation-based techniques \citep{sasaki2019transfer, encinar2019logarithmic}. This work focuses on the use of the aforementioned linear coherence, which is a transparent and widely applied input-output system analysis technique. With an ideal actuator where the energy in the estimated off-wall signal is perfectly nullified, only the fraction of turbulent fluctuating energy at the output location, quantified by the LCS, is eliminated. Hence, this work uses the LCS to explore the feasibility of opposition control in the boundary layer, based on a wall-based input sensor. To this end, we employ the LCS of $u$- and $w$-fluctuations away from the wall with $u_\tau$-fluctuations (streamwise friction velocity fluctuations) together with the energy spectra from temporal and spatial DNS data at $Re_\tau \approx 590$ and $2000$. Although numerical simulations suggest that manipulation of the QSVs with control schemes utilizing both the wall-normal and spanwise velocity fluctuations at some off-wall location as input signals yield higher drag reduction than those with sensors measuring streamwise velocity \citep{choi1994active}, we chose $u_\tau$-fluctuations as the input for this study since these can be measured more reliably in practical high-Reynolds-number flows.

Throughout this study $x,y$ and $z$ are the streamwise, spanwise and wall-normal directions, with $u$, $v$ and $w$ representing the respective fluctuating velocity components. Capitalisation and angled brackets, $\langle \rangle$, show averaged quantities, while lower cases correspond to fluctuations from the time-averaged mean values. The superscript `$+$' denotes viscous scaling of velocity (e.g. $U^+ = U/U_\tau$) and length (e.g. $z^+ = z U_\tau / \nu$) and a `$\sim$' over a letter indicates the Fourier transform.

\section{Numerical data of wall-bounded turbulence}
\begin{table}
\centering
\begin{tabular}{ c c c c c c }
 Simulation data & $Re_\tau$ & Spatial & Temporal & $\Delta y^+$ & $\Delta t^+$ \\
 			\midrule 
DNS (current study) & $h U_\tau/\nu \approx 590$ & \checkmark & \checkmark & 3.6 & 0.89 \\  
DNS \citep{sillero2013one} & $\delta_{99} U_\tau/\nu \approx 2000$ & \checkmark & & 3.7 & - \\
LES \citep{eitel2014simulation} & $\delta_{99} U_\tau/\nu \approx 2500$  & & \checkmark & 8 & 0.47 
\end{tabular}
	\caption{Details of the numerical data sets.}
	\label{table:details}
\end{table}
The first data set  was generated at the University of Melbourne, using a fully-conservative fourth-order finite difference code. The code has been verified in previous DNS studies of wall-bounded flows~\citep{chung2014,chung2015}. The computational domain is open-channel flow, with a domain size of $2\pi h \times \pi h \times h$ in the streamwise, spanwise and wall-normal directions, respectively, where $h$ is the open-channel height. Periodic boundary conditions are applied in the streamwise and spanwise directions, and no-slip and free-slip boundary conditions are applied at the bottom and top boundaries, respectively. The flow was driven by a constant pressure gradient, adjusted such that $Re_\tau = U_\tau h/\nu = 590$. The grid resolutions are $\Delta x^+ = 7.2$ (streamwise) and $\Delta y^+ = 3.6$ (spanwise), which are fine enough for DNS. Note that `+' signifies viscous scaling by the mean friction velocity $U_\tau$ and kinematic viscosity $\nu$. The data were recorded every $\Delta t^+ = 0.89$ for a duration of $ T \approx 0.99 h/U_\tau$, after the simulation reaches the statistically stationary state.

The second data set is of a TBL by \citet{sillero2013one}, from which streamwise--spanwise planes of data were extracted. These planes extend $\sim 12 \delta_{99}$ in the streamwise direction with $Re_\tau=\delta_{99} U_\tau/\nu \approx 2000$ at the streamwise center of the planes. This streamwise extension ensures inclusion of large scale turbulent structures while maintaining an acceptable Reynolds number variation between $Re_\theta=\theta U_\infty/\nu \approx 5110$ and 6010 at the upstream and downstream ends of the planes, where $\theta$ is the momentum thickness. The spanwise resolution is $\Delta y^+=3.7$.

Finally, a third data set comprises time-series of $u$-velocity fluctuations from a finely resolved large-eddy simulation (LES) of TBL flow by \cite{eitel2014simulation}. Simultaneous time-series at a fixed streamwise location, spanning the entire boundary layer in $z$ and $y$ directions were extracted. The Reynolds number at the $x$ location where the data are collected is $Re_\tau=\delta_{99} U_\tau/\nu \approx 2500$. Note that these data resemble simultaneous data collected by a 2D wall-normal/spanwise grid of hot-wires in a wind tunnel. The data were recorded every $\Delta t^+=0.47$ for a duration of $T \approx 3.8 \delta_{99}/U_\tau$, and the spanwise resolution is $\Delta y^+=8$. Details of the numerical data are summarized in table 1.

\begin{figure}
\includegraphics[width=1.00\textwidth]{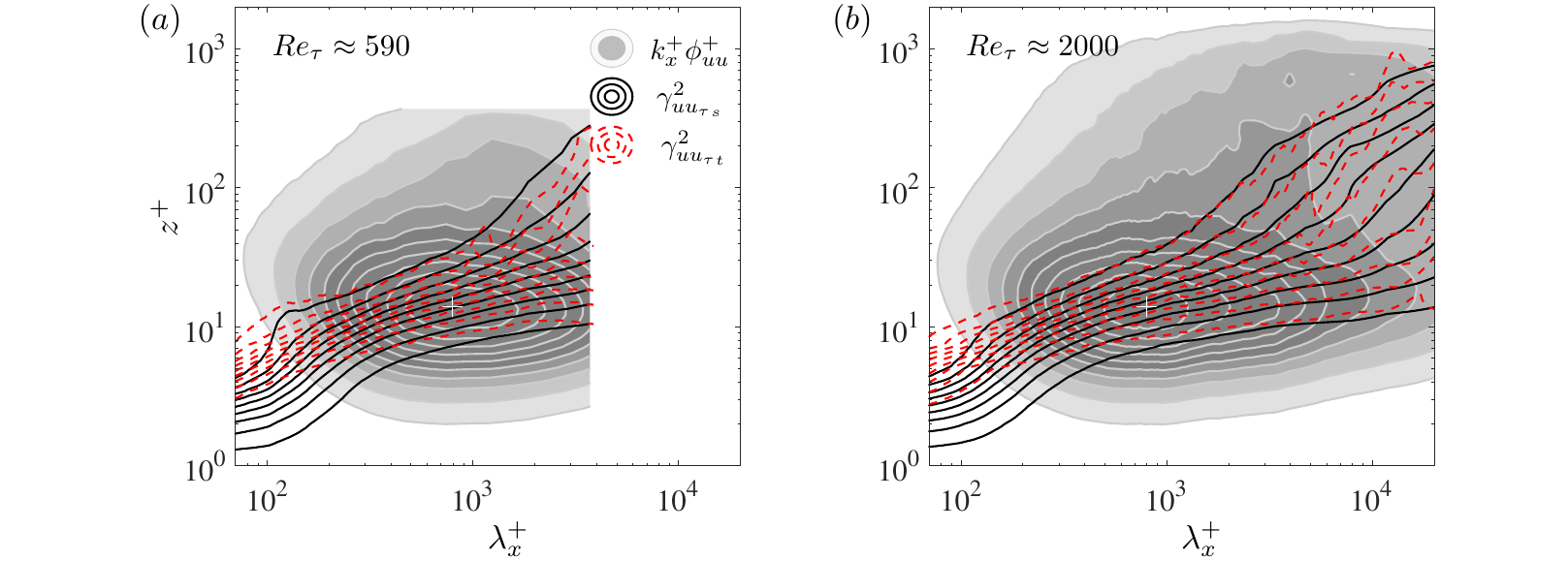}
\vspace*{-0.7cm}
\caption{Coherence and energy spectrograms. $(a)$ $Re_\tau \approx 590$ $(b)$ $Re_\tau \approx 2000$. Filled contours, $k_x \phi_{uu}/U_\tau^2$ (levels: 0.2:0.2:2.2); solid lines, $\gamma_{{uu_\tau}_s}^2$ computed from spatial velocity fluctuations (levels: 0.1:0.1:0.9); dashed lines, $\gamma_{{uu_\tau}_t}^2$ calculated from temporal velocity fluctuations together with Taylor's turbulence hypothesis (levels: 0.1:0.1:0.9).}
\label{fig:coherence_energy_spec}
\end{figure}

\section{Coherence of near wall region turbulence}

\subsection{Spatial versus temporal coherence spectrograms}
\begin{figure}
\includegraphics[width=1.00\textwidth]{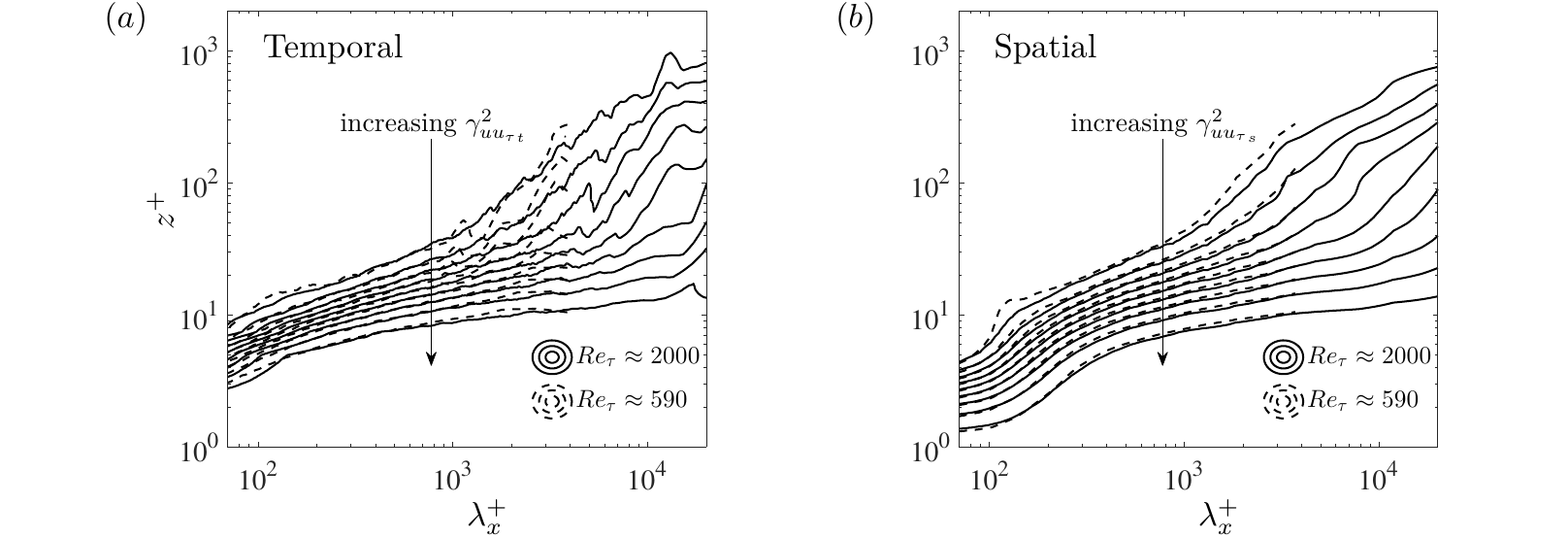}
\vspace*{-0.7cm}
\caption{Reynolds number comparison of coherence spectrograms (levels: 0.1:0.1:0.9). $(a)$ Temporal spectrograms $\gamma_{{uu_\tau}_t}^2$. $(b)$ Spatial spectrograms $\gamma_{{uu_\tau}_s}^2$. Dashed lines, $Re_\tau \approx 590$; solid lines, $Re_\tau \approx 2000$. The arrow direction indicates increase in the coherence spectra levels.}
	\label{fig:Re_comparison}
\end{figure}

In typical experimental control efforts, temporal data acquired by a stationary wall-based sensor forms the input. Recently, \citet{baars2017self} noted differences between coherence spectra obtained from high Reynolds number experimental (temporal) and moderate Reynolds number DNS (spatial) data of streamwise velocity fluctuations ($u$) in the near wall region ($z^+<80$). Their results showed that the coherence between the $u$-fluctuations at a point very close to the wall and those away from the wall vanished in the experimental spectrogram for approximately $\lambda_x^+<7000$ and $z^+<80$, while the DNS spectrogram showed noticeable coherence in that ($\lambda_x,z$)-domain. Since the energy associated with the QSVs resides in that domain of the energy spectrogram (with its peak at $\lambda_x^+ \approx 800$ and $z^+ \approx 15$), the aforementioned discrepancy has implications for the practical feasibility of wall-based opposition control of the near wall cycle. In \citet{baars2017self} it was hypothesized that the absence of coherence was caused by using temporal data, rather than spatial data for computing the LCS. We here explore the issue of spatial versus temporal coherence by comparing $\gamma_{uu_\tau}^2$ spectra from temporal and spatial data at $Re_\tau \approx 590$ and $\approx 2000$ for $\Delta x= \Delta y=0$. Figures \ref{fig:coherence_energy_spec}(a) and \ref{fig:coherence_energy_spec}(b) indicate $\gamma_{{uu_\tau}_s}^2 (z^+;\lambda_x^+)$ (spatial coherence) and $\gamma_{{uu_\tau}_t}^2 (z^+; \lambda_x^+)$ (temporal coherence) overlaid on the viscous-scaled premultiplied energy spectrogram $k_x \phi_{uu}/U_\tau^2$ at $Re_\tau \approx 590$ and $\approx 2000$, respectively. Here $k_x=2 \pi/\lambda_x$ is the streamwise wavenumber. In figure \ref{fig:coherence_energy_spec}(b), DNS data of \citet{sillero2013one} and LES data of \citet{eitel2014simulation} are used to calculate $\gamma_{{uu_\tau}_s}^2$ and $\gamma_{{uu_\tau}_t}^2$, respectively. Firstly, contrary to the observation of \cite{baars2017self} from temporal hot-wire data, $\gamma_{{uu_\tau}_t}^2$ is non-zero for $z^+<100$ and $\lambda_x^+<7000$. Secondly, $\gamma_{{uu_\tau}_t}^2 \approx \gamma_{{uu_\tau}_s}^2$ for $\lambda_x^+ > 500$ covering the near wall energy site in the energy spectrogram (indicated by a `+' marker) corresponding to the QSVs. This implies that the use of temporal data from a stationary wall-sensor, as input for off-wall opposition control of near wall turbulence, is as effective as using spatial data. It is also noted that the temporal and spatial coherence spectra differ significantly in the region $z^+<15$ and $\lambda_x^+<500$, which is presumably due to the mismatch between the local mean and the true structure convection velocity in that region \citep{del2009estimation, drozdz2017amplitude, liu2020input}. Figures \ref{fig:Re_comparison}(a) and \ref{fig:Re_comparison}(b) compare the $\gamma_{{uu_\tau}_t}^2$ and $\gamma_{{uu_\tau}_s}^2$ spectrograms at $Re_\tau \approx 590$ and $\approx 2000$, respectively. Evidently, both temporal and spatial coherence spectra are Reynolds number independent when plotted against viscous-scaled variables for the small to moderate wavelengths ($\lambda_x<1000$).

\begin{figure}
\includegraphics[width=1.00\textwidth]{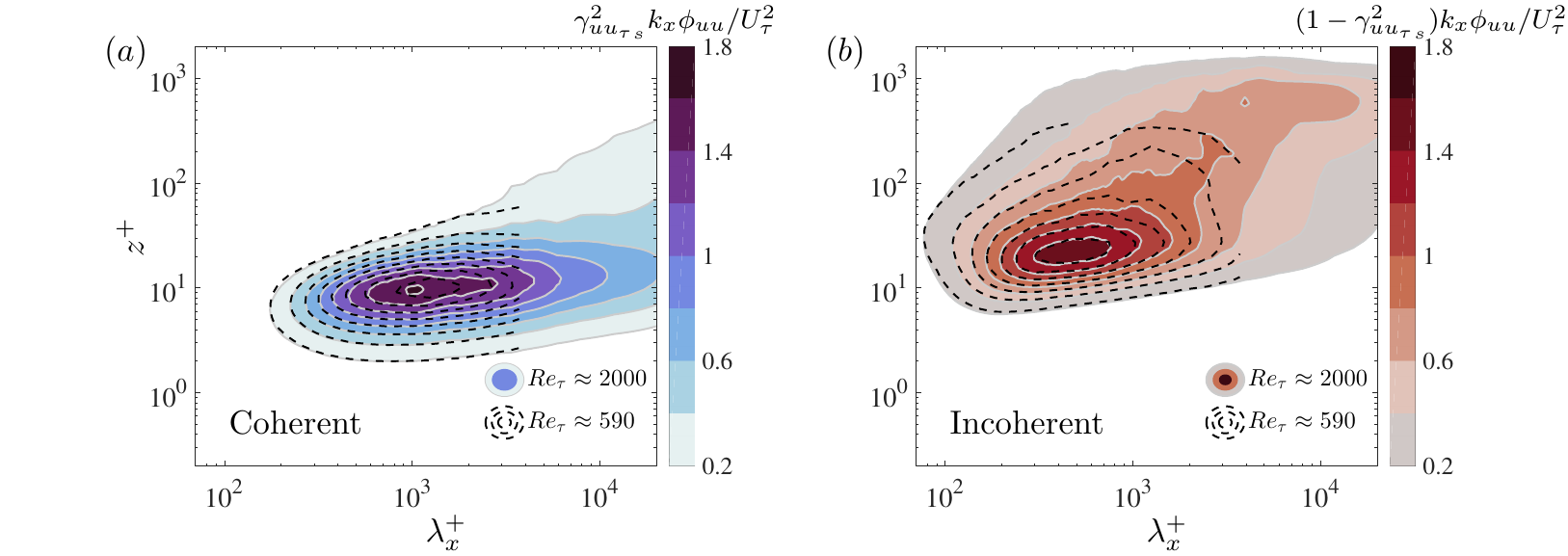}
\vspace*{-0.7cm}
\caption{Reynolds number comparison of wall-coherent and wall-incoherent portions of the pre-multiplied energy spectrograms (levels: 0.2:0.2:1.6). $(a)$ Coherent portion, $\gamma_{{uu_\tau}_s}^2 k_x \phi_{uu}/U_\tau^2$. $(b)$ Incoherent portion, $(1-\gamma_{{uu_\tau}_s}^2) k_x \phi_{uu}/U_\tau^2$. Dashed line contours, $Re_\tau \approx 590$; filled contours, $Re_\tau \approx 2000$. Spatial data are used to computed these correlation spectra.}
	\label{fig:energy_decomposition}
\end{figure}

\subsection{Coherent and incoherent portions of the streamwise turbulence intensity}
Through the use of the LCS it is possible to determine the fractions of the streamwise energy spectrograms that are coherent ($\gamma_{{uu_\tau}}^2 k_x \phi_{uu}/U_\tau^2$) and incoherent ($[1-\gamma_{{uu_\tau}}^2] k_x \phi_{uu}/U_\tau^2$) relative to the wall-based signal used in creating the LCS \citep[the coherent portion reflects the energy that could be predicted via a linear stochastic estimation procedure, when taking the wall-signal as input;][]{adrian1979conditional}. The wall-coherent and wall-incoherent spectrograms are shown in figures \ref{fig:energy_decomposition}(a) and \ref{fig:energy_decomposition}(b), respectively, at $Re_\tau \approx 590$ (dashed line contours) and $Re \approx 2000$ (filled contours). Here $\Delta x=\Delta y=0$. The coherent portions of the $u$-spectrograms collapse for $\lambda_x^+<1000$. Beyond this, an increasing amount of large wavelength energy at all wall distances with increasing $Re_\tau$ is observed, as is expected due to the growing energetic range of viscous-scaled wavelengths with increasing $Re_\tau$. It is also noticeable from figure \ref{fig:energy_decomposition} (b), that there is substantial incoherent energy quite close to the wall, and even at scales and wall-normal locations that we would typically associate with the near-wall structures. The wall-incoherent spectrograms collapse for $z^+<20$, and $\lambda_x^+ < 2000$. It will be shown that the extra large scale wall-incoherent energy (residing over the wavelength domain $\lambda_x^+ > 2000$) at higher $Re_\tau$ does not have significant contribution to the streamwise turbulence intensity. We can integrate the coherent and incoherent spectrograms over all wavelengths to obtain the coherent and incoherent streamwise turbulence intensity $\overline{u^2}^+_C$ and $\overline{u^2}^+_{IC}$, respectively, following
\begin{eqnarray}
 \label{eq:coh}
 \overline{u^2}^+_C\left(z\right) &=& \int_{0}^{\infty} \gamma_{{uu_\tau}}^2\left(z;\lambda_x\right) k^+_x \phi^+_{uu}\left(z;\lambda_x\right) {\rm d}\log(\lambda_x), \\ 
 \label{eq:incoh} 
 \overline{u^2}^+_{IC}\left(z\right) &=& \int_{0}^{\infty} \left[1-\gamma_{{uu_\tau}}^2\left(z;\lambda_x\right)\right] k^+_x \phi^+_{uu}\left(z;\lambda_x\right) {\rm d}\log(\lambda_x).
\end{eqnarray}
with the total turbulence intensity being the sum of both, $\overline{u^2}^+\left(z\right) = \overline{u^2}^+_C\left(z\right)+\overline{u^2}^+_{IC}\left(z\right)$.

\begin{figure}
\includegraphics[width=1.00\textwidth]{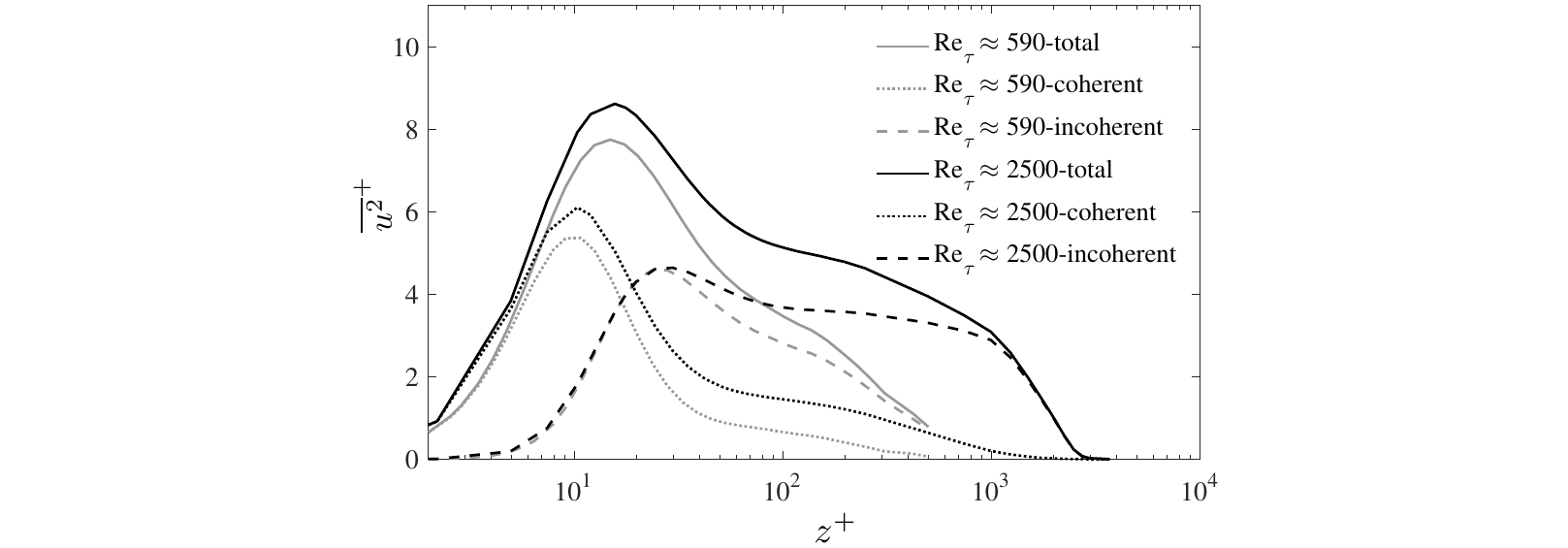}
\vspace*{-0.7cm}
\caption{Decomposition of turbulence intensity profiles into contributions that are stochastically coherent with the wall via (\ref{eq:coh}) and stochastically wall-incoherent via (\ref{eq:incoh}). Two sets of profiles are shown for $Re_\tau \approx 590$ and $Re_\tau \approx 2500$.}
	\label{fig:var_coherence}
\end{figure}

\begin{figure}
\includegraphics[width=1.00\textwidth]{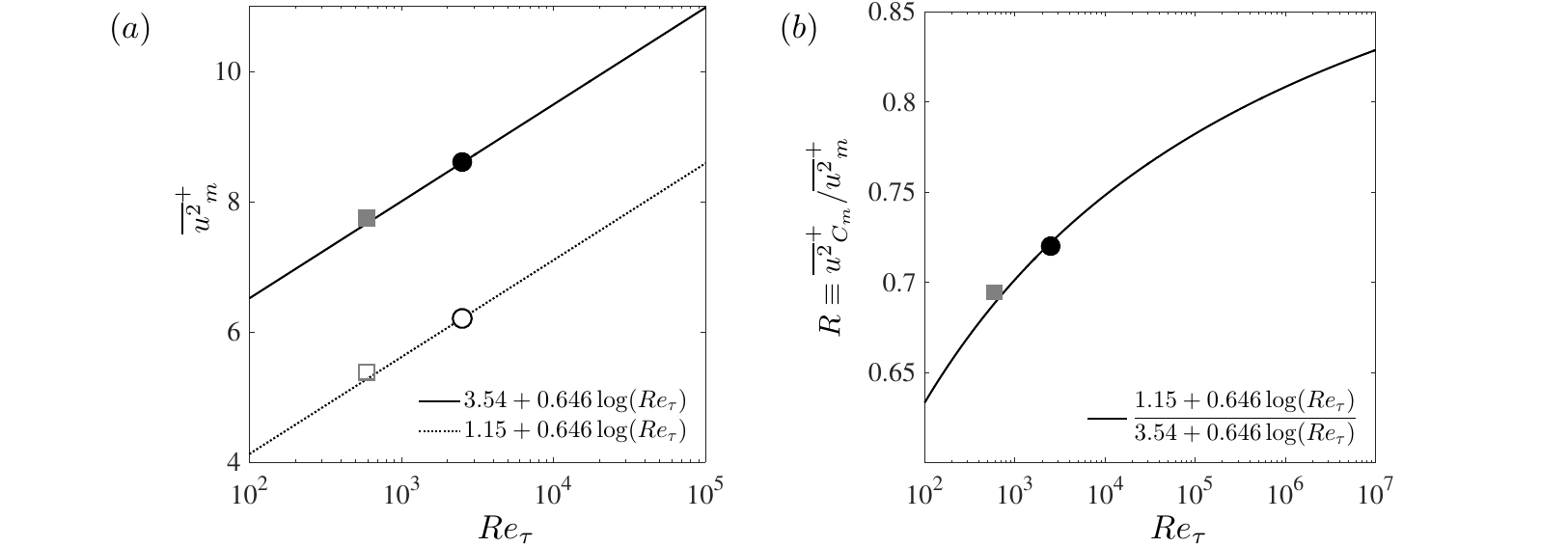}
\vspace*{-0.7cm}
\caption{(a) Turbulence intensity peak of the total intensity profile (solid line) and coherent intensity profile (dotted line). Solid and open symbols correspond to the peaks in total intensity and coherent intensity profiles (at $Re_\tau \approx$ 590 and 2500), respectively. (b) Ratio of the peak in coherent intensity profile to the peak in total intensity profile. }
	\label{fig:var_coherence_Re}
\end{figure}

Profiles of the wall-coherent and wall-incoherent contributions to $\overline{u^2}$ are shown in figure~\ref{fig:var_coherence} for $Re_\tau \approx 590$ and 2500. Here temporal data are used to calculate the turbulence intensities, since the spatial energy spectra exclude energy of wavelengths beyond the limited size of the computational domain resulting in attenuated turbulence intensities obtained from the integration. Figure \ref{fig:var_coherence} illustrates that the incoherent intensity profiles collapse for $z^+ \leq 20$, suggesting that only the motions that are coherent with the wall contribute to the peak growth of $\overline{u^2}$ with $Re_\tau$ \citep{marusic2017scaling, samie2018fully}. While the peak in $\overline{u^2}$ is at $z^+ \approx 15$, the ${\overline{u^2}^+_C}$ profiles peak at a slightly smaller wall-normal location of $z^+ \approx 10$. These peak values for $Re_\tau \approx 590$ and 2500 are shown in figure \ref{fig:var_coherence_Re}(a) against $Re_\tau$, together with the log-linear relation from \citet{samie2018fully} for the peak of $\overline{{u^2}}^+$ and a log-linear relation for the peak of $\overline{{u^2}}^+_{C}$. \cite{samie2018fully} extracted the former relation from their fully-resolved experimental data up to $Re_\tau \approx \num{20000}$. Assuming that the incoherent portion of the turbulence intensity for $z^+<20$ is independent of $Re_\tau$ (as seen for $Re_\tau \approx 590$ and 2500), a relation with the same slope can be deduced for the peak of ${\overline{u^2}^+_C}$. The ratio of these peaks ($R \equiv {\overline{{u^2}}^+_{C_m}}/{\overline{{u^2}}^+_{m}}$) increases with $Re_\tau$ (figure \ref{fig:var_coherence_Re}b), due to the growing range of energetic large-scales that are wall-coherent.

The ratio $R$ indicates the fraction of the fluctuating energy of $u$ at $z^+=15$ (integrated over all scales) that can be predicted correctly using a linear stochastic estimate or Kalman filter-type of approach based on measurements of $u$ made at the wall. For instance, when $R \approx 0.7$ at $Re_\tau = 10^3$, only 70\% of the energy in the $u$-fluctuations could be estimated and be targeted with opposition control (in a stochastic sense). Even if actuators could perfectly nullify those estimated fluctuations, 30\% of the energy remains unaffected. The increasing trend of $R$ with $Re_\tau$ shows that the performance of a controller targeting all scales of motion in the near wall region can improve with increasing $Re_\tau$.

On the other hand, the energy associated with the near-wall wall-coherent streaks (small to medium wavelengths) form an increasingly smaller component of the wall-coherent energy as Reynolds number grows (due to the growth of the large scale portion of the spectrogram with $Re_\tau$ while the small to medium scale portion is constant in the viscous-scaled spectrogram as shown in figure \ref{fig:energy_decomposition}a). 

\begin{figure}
\includegraphics[width=1.00\textwidth]{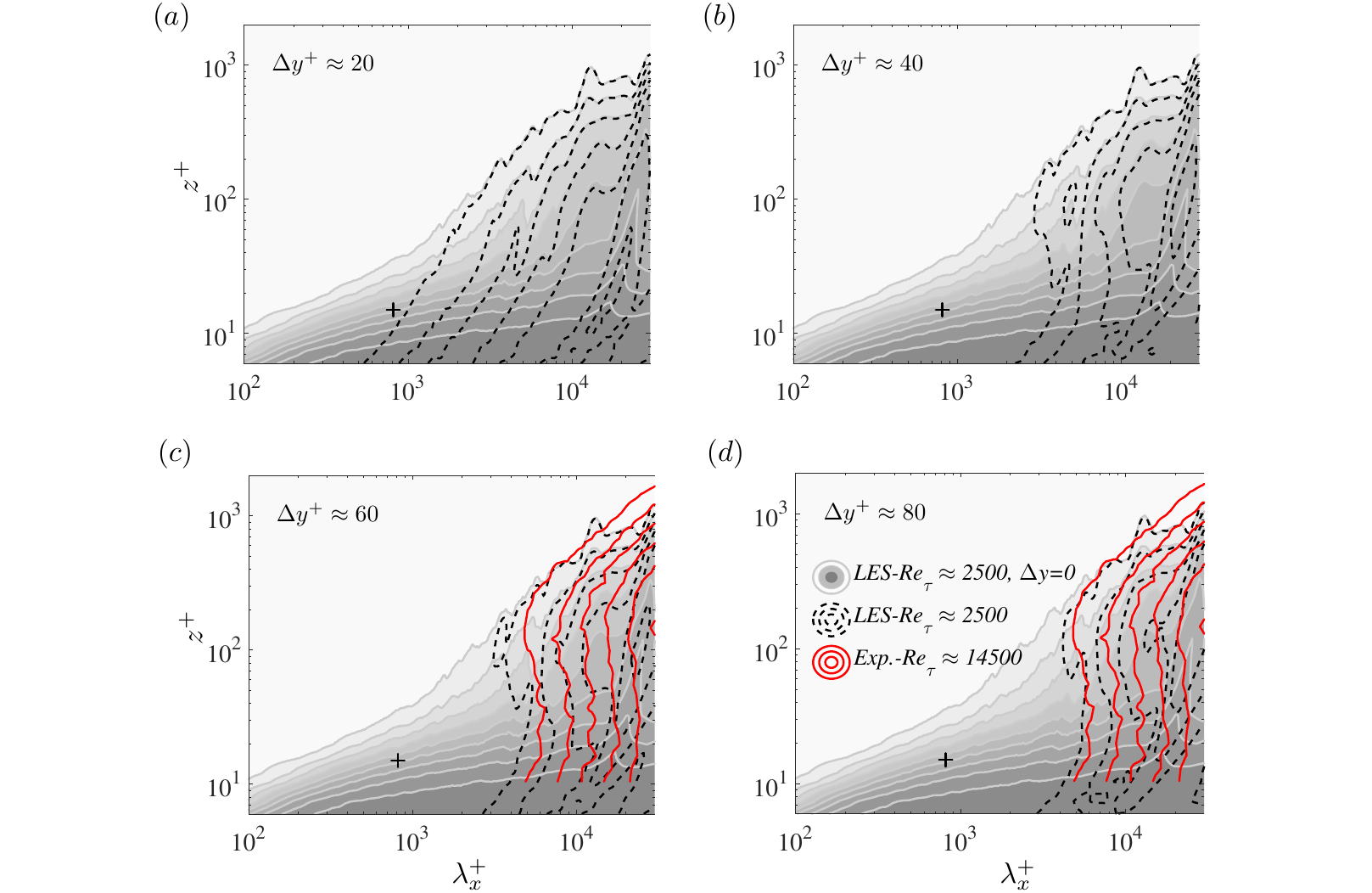}
\vspace*{-0.75cm}
	\caption{Effect of spanwise distance between sensor and actuator ($\Delta y^+$) on the temporal coherence spectrograms $\gamma_{{uu_\tau}_t}^2$. Filled contours, LES coherence spectrogram at $Re_\tau \approx 2500$ and $\Delta y=0$; dashed line contours, LES coherence spectrogram at $Re_\tau \approx 2500$ and varying $\Delta y$; solid red line contours, experimental coherence spectrogram at $Re_\tau=14500$. ($a$) $\Delta y^+ = 20$; ($b$) $\Delta y^+ = 40$; ($c$) $\Delta y^+ = 60$; ($d$) $\Delta y^+ = 80$. }
\label{fig:offset_effect}
\end{figure}

\subsection{Effect of spanwise distance between sensor and actuator}

The typical diameter of the QSVs is in the order of $d^+=20-50$ \citep{kim1987turbulence}. Because a significant portion of the wall-coherent energy is driven by the QSV-type structures, its typical length-scale in relation to a spanwise offset between sensors and actuators has a major influence on the current results. The LCS can be used to explore the effect of spanwise distance between the sensor and the actuator as shown in figure \ref{fig:offset_effect}, where $\gamma_{{uu_\tau}}^2$ is presented for $\Delta y^+=20$, 40, 60 and 80 for $Re_\tau\approx 2500$. To isolate the effect of spanwsie distance on the $\gamma_{{uu_\tau}}^2$ spectra, the streamwise separation is taken as $\Delta x=0$. The near wall peak in the $u$-energy spectrogram is shown with `+' at $z^+=15$ and $\lambda_x^+ =800$. It is evident that for a spanwise distance as small as $\Delta y^+=20$, the coherence between the velocity signals drops to zero at $(z^+,\lambda_x^+)=(15,800)$. This has the implication that a velocity signal in the near wall region cannot be targeted upon correctly, given a controller's input wall-signal, if the spanwise distance between the sensor (input signal) and  actuator (targeted signal) is not carefully controlled. As a guideline, the spanwise misalignment for sensor--actuator pairs must be maintained below $20\nu/U_\tau$. This is smaller than the lower limit of the typical diameter of the QSVs. For a TBL airflow under standard temperature and pressure conditions, with a boundary layer thickness of $\delta=1$\,m and $Re_\tau=10^5$, this translates to $\Delta y=0.2$\,mm. Such tight tolerances for spanwise alignment might pose significant challenges for the hardware of active, wall-based opposition control systems. Also shown in figures~\ref{fig:offset_effect}(c)-(d) is the experimental $\gamma_{{uu_\tau}}^2$ spectrogram of \citep{baars2017self} at $Re_\tau=14500$. The resemblance between the experimental $\gamma_{{uu_\tau}}^2$ spectra and LES $\gamma_{{uu_\tau}}^2$ spectra in figure \ref{fig:offset_effect}(d) suggests that the vanishing small-scale coherence in the near wall region reported by \citet{baars2017self} might be caused by a spanwise misalignment of the sensors in the experiment. 

\begin{figure}
\includegraphics[width=1.00\textwidth]{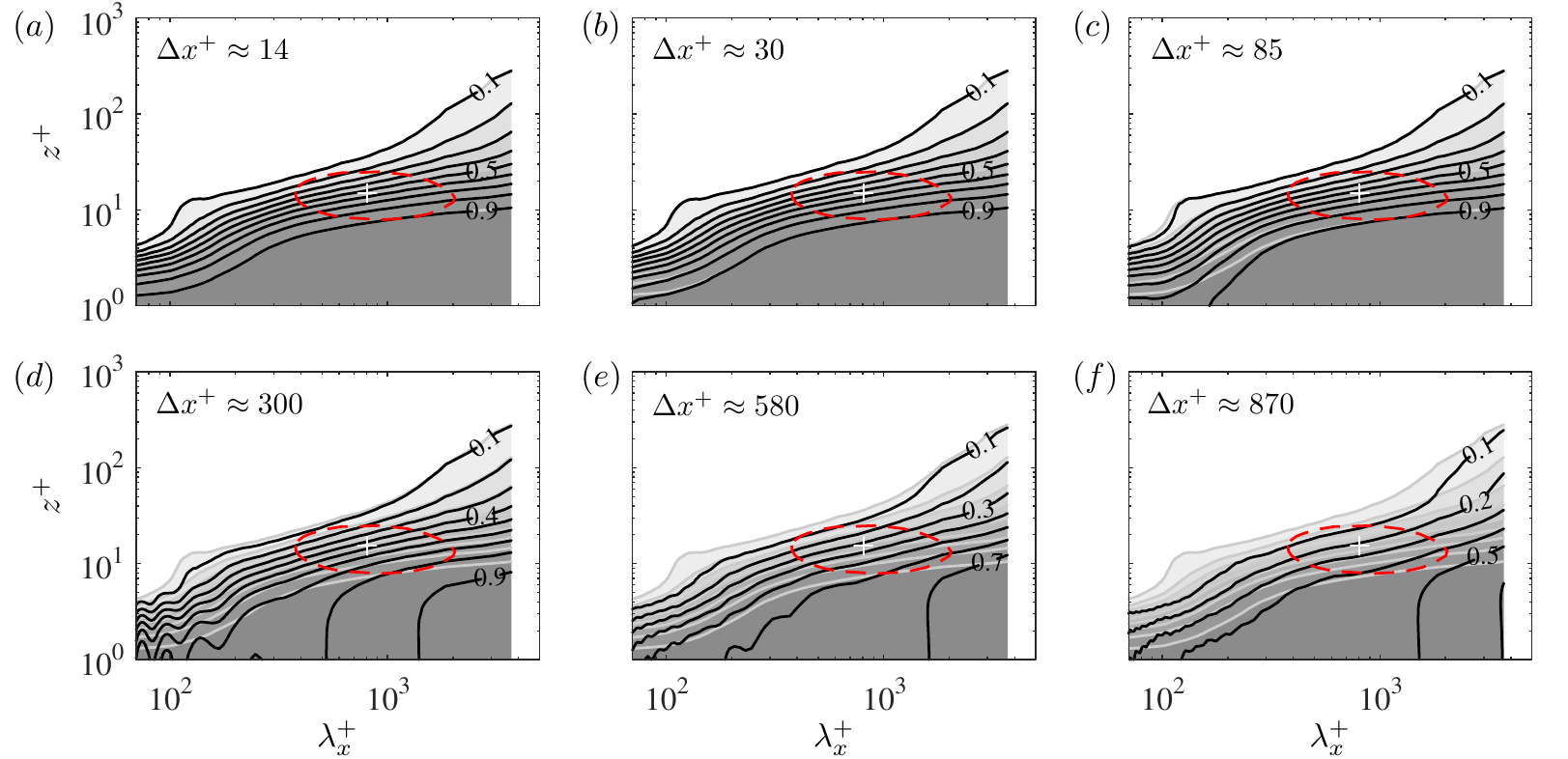}
\vspace*{-0.75cm}
\caption{Effect of streamwise distance between wall reference sensor and actuation points ($\Delta x$) on the spatial coherence spectrogram $\gamma_{{uu_\tau}_s}^2$ at $Re_\tau \approx 590$. The filled contours show the coherence spectrogram for $\Delta x^+ = 0$ (levels: 0.1:0.1:0.9), while the contour lines show coherence spectrograms for varying $\Delta x^+$. Dashed lines show $u$-energy spectra contour level $k_x^+ \phi_{uu}^+=1.8$ and `+' markers show the peak in the energy spectrogram. ($a$) $\Delta x^+ = 14$; ($b$) $\Delta x^+ = 30$; ($c$) $\Delta x^+ = 85$; ($d$) $\Delta x^+ = 300$; ($e$) $\Delta x^+ = 580$; ($f$) $\Delta x^+ = 870$. }
	\label{fig:streamwise_offset_a-f}
\end{figure}

\subsection{Effect of streamwise distance between sensor and actuator}
\begin{figure}
\includegraphics[width=1.00\textwidth]{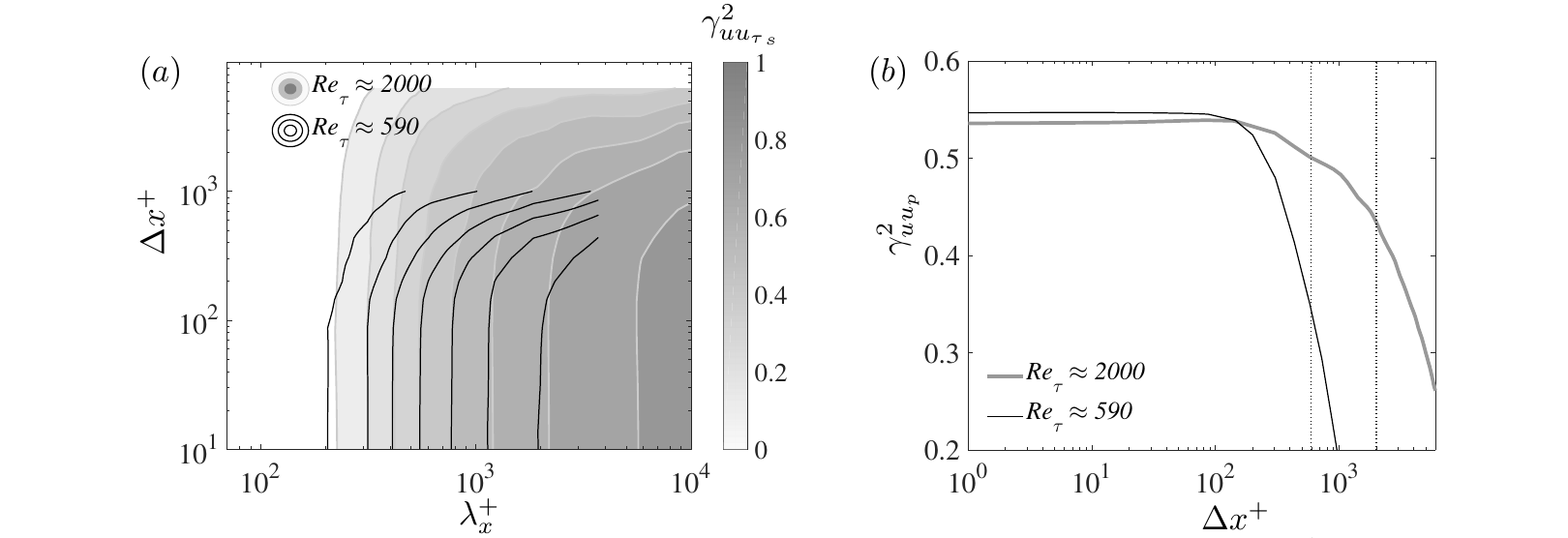}
\vspace*{-0.75cm}
	\caption{ ($a$) Reynolds number effect on the spatial coherence spectrogram at $z^+=15$ as a function of $\lambda_x^+$ and $\Delta x^+$. Contour lines are associated with $Re_\tau \approx 590$ while filled contours correspond to $Re_\tau \approx 2000$.  Contour levels start from 0.1 in steps of 0.1.($b$) Coherence magnitude at the near wall energy site, i.e. $z^+=15$ and $\lambda_x^+=810$ as a function of $\Delta x^+$. Vertical dotted lines show $\Delta x^+= \delta^+$ for $Re_\tau \approx 590$ and 2000.  }
	\label{fig:gamma_dx_lambda}
\end{figure}

In real closed-loop control systems, the sensor and the actuator are often placed at a finite streamwise distance ($\Delta x$) from each other to account for the controller's processing time and the delay associated with the actuator's response \citep[see e.g.][]{rathnasingham2003active, rebbeck2006wind}. This streamwise distance leads to a lower coherence between the velocity measured by the sensor and that passing over the actuator. Figure \ref{fig:streamwise_offset_a-f} shows the $\gamma_{{uu_\tau}}^2$ spectra at $Re_\tau \approx 590$ for various sensor--actuator streamwise distances (line contours) overlaid on the $\gamma_{{uu_\tau}}^2$ spectra for zero sensor--actuator distance (filled contours). To isolate the effect of streamwise distance on the $\gamma_{{uu_\tau}}^2$ spectra, the spanwise distance is kept zero ($\Delta y=0$). The near wall peak in the energy spectra, which is related to the QSVs and near wall energetic cycle of streaks, is shown with `+' symbols, and the $u$-energy spectra contour level $k_x^+ \phi_{uu}^+=1.8$ is shown with dashed lines. By focusing on $\gamma_{{uu_\tau}}^2$ near this peak, one can see that, up to $\Delta x^+ \approx 85$, the $\gamma_{{uu_\tau}}^2$ spectra remain virtually unchanged. However, with further increase in $\Delta x^+$ the coherence around the near wall peak starts to diminish. Similar results are obtained for $Re_\tau \approx 2000$ (not shown here for brevity). Figure \ref{fig:gamma_dx_lambda}(a) plots the coherence spectra at the approximate location of the near-wall peak ($z^+ = 15$) as a function of the streamwise separation distance $\Delta x^+$ and the streamwise wavelength $\lambda_x^+$; i.e. $\gamma_{{uu_\tau}}^2(z^+=15, \Delta x^+; \lambda_x^+)$ for $Re_\tau \approx 590$ and 2000. Note that the large wavelength portion of the coherence spectra ($\lambda_x^+>3500$) at $Re_\tau \approx 590$ is missing due to the limited size of the computational domain. Figure \ref{fig:gamma_dx_lambda}(b) shows curves of $\gamma_{{uu_\tau}}^2$ at $z^+=15$ and $\lambda_x^+=810$ for these Reynolds numbers. It appears that, $\gamma_{{uu_\tau}}^2$ at $z^+=15$ is independent of $\Delta x^+$ for small $\Delta x^+$. Also dependence on $Re_\tau$ is weak. Moreover, the $\Delta x^+$ range in which $\gamma_{{uu_\tau}}^2$ remains independent of $\Delta x^+$ expands with $Re_\tau$. More data over a range of $Re_\tau$ are required to determine an explicit formulation for this $Re_\tau$ dependency. According to figure \ref{fig:gamma_dx_lambda}(b), $\gamma_{{uu_\tau}}^2 \approx 0.55$ at $(z^+,\lambda_x^+)=(15,810)$ (i.e. at the near wall peak location) for $\Delta x^+=0$ and remains constant with increasing $\Delta x^+$ up to a limit, which appears to be an increasing function of Reynolds number. The coherence $\gamma_{{uu_\tau}}^2$ decreases rapidly with further increase in $\Delta x^+$ after that limit and drops to $90 \%$ of its maximum value at $\Delta x^+ \approx 250$ and 950 for $Re_\tau=590$ and 2000, respectively. The fact that the highest $\gamma_{{uu_\tau}}^2$ value is only 0.55 at the near wall peak location suggests that the closed-loop drag reduction schemes using wall-based sensing of skin friction fluctuations to target QSVs can only achieve limited success. The wall-coherence $\gamma^2_{u u_\tau}$ at $z^+ = 15$ is much higher for longer streamwise structures (larger $\lambda_x$), which might suggest that opposition control schemes targeting a wider range of scales of motion could be a viable avenue for future control schemes with wall-based sensing (notwithstanding the increasing challenges this then poses for wall-based actuation).

\subsection{Coherence of wall-normal velocity component with the streamwise friction velocity}

\begin{figure}
\includegraphics[width=1.00\textwidth]{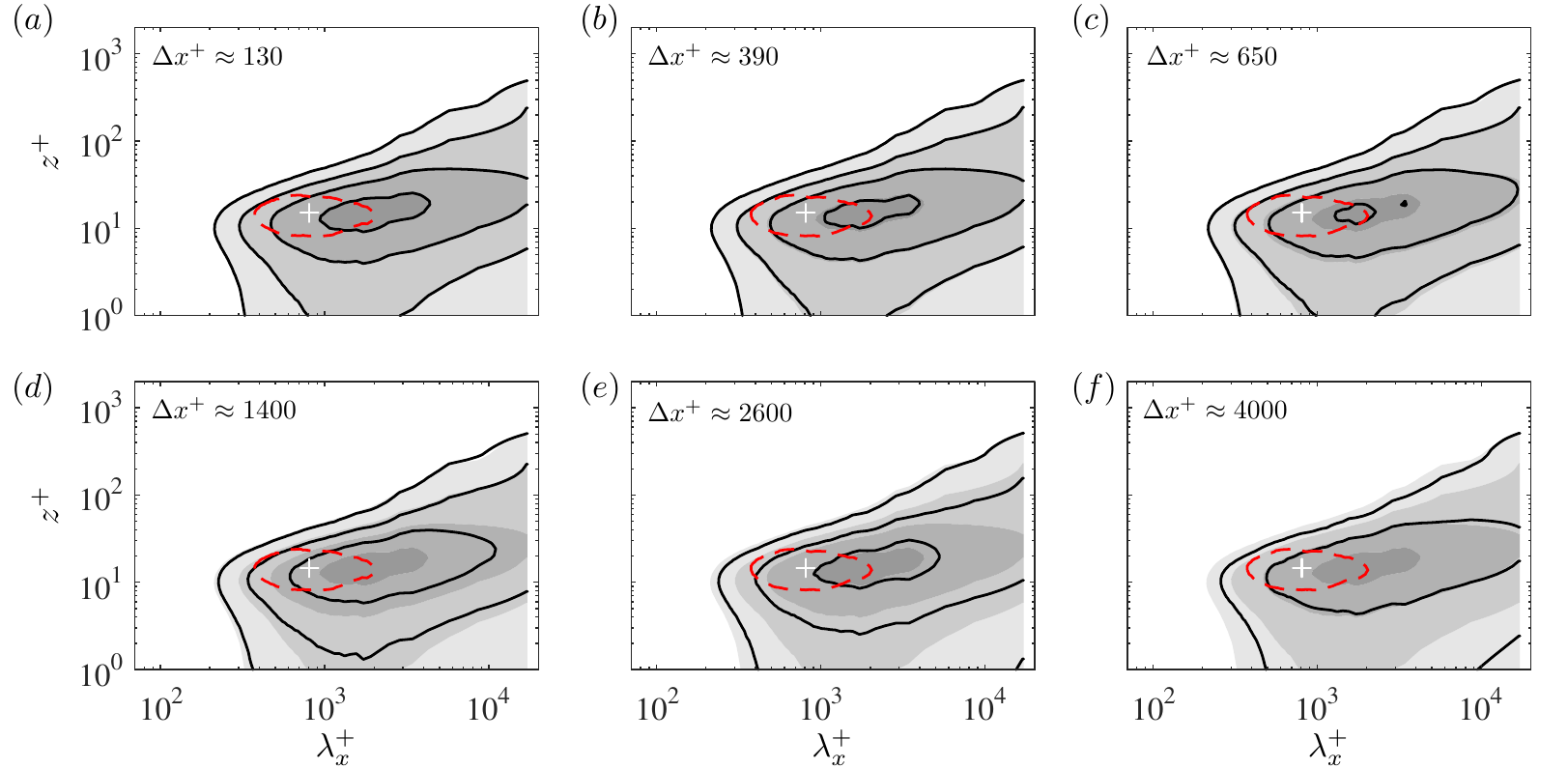}
\vspace*{-0.75cm}
	\caption{Effect of streamwise distance between wall reference sensor and actuation points ($\Delta x$) on the spatial wall-normal coherence spectrogram $\gamma_{{wu_\tau}_s}^2$ at $Re_\tau \approx 2000$. The filled contours show the coherence spectrogram for $\Delta x^+ = 0$ (levels: 0.1:0.1:0.4), while the contour lines show coherence spectrograms for varying $\Delta x^+$. The dashed lines show premultiplied energy spectrogram contour level $k_x^+ \phi_{uu}^+=1.8$ and `+' markers show the peak in the energy spectrogram.}
	\label{fig:w_streamwise_offset_a-f}
\end{figure}

\begin{figure}
\includegraphics[width=1.00\textwidth]{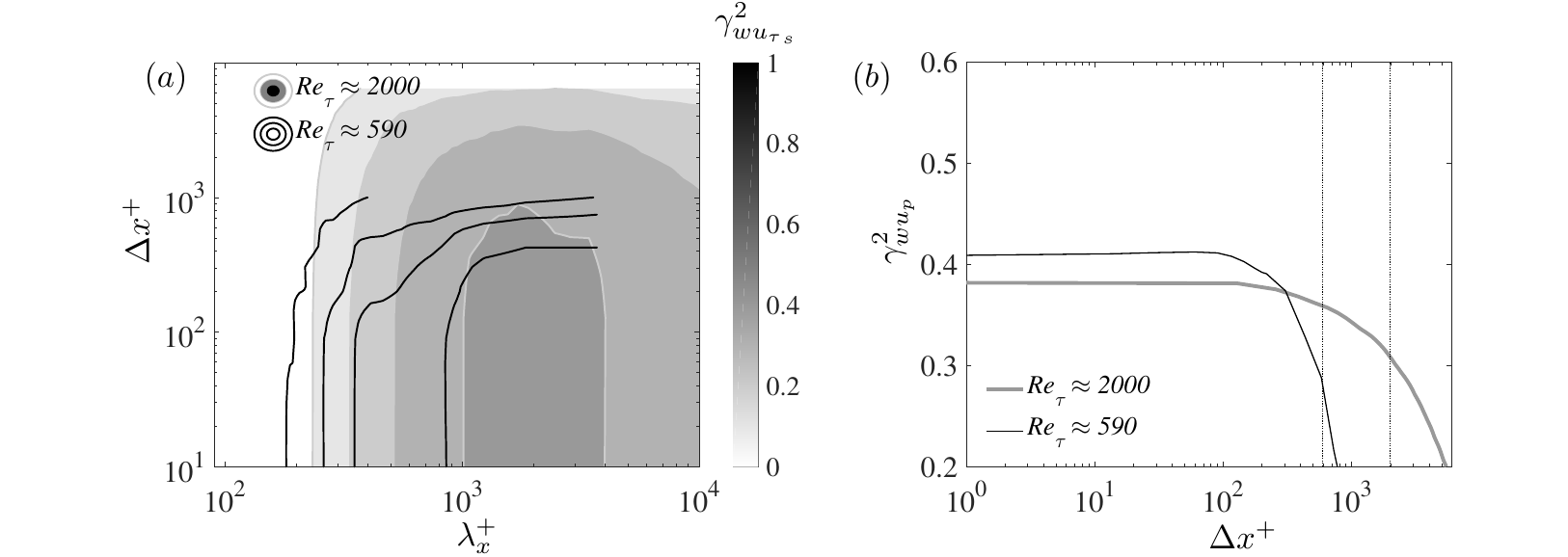}
\vspace*{-0.75cm}
	\caption{ ($a$) Wall-normal coherence spectrogram at $z^+=15$ as a function of $\lambda_x^+$ and $\Delta x^+$. Contour lines are associated with $Re_\tau \approx 590$ while filled contours correspond to $Re_\tau \approx 2000$. Contour levels start from 0.1 in steps of 0.1. ($b$) Wall-normal coherence magnitude at the near wall energy site, i.e. $z^+=15$ and $\lambda_x^+=810$ as a function of $\Delta x^+$. Vertical dotted line shows $\Delta x^+= \delta^+$.}
	\label{fig:w_gamma_dx_lambda}
\end{figure}

We have focused thus far on the coherence of $u$-fluctuations at some off-wall locations with friction velocity fluctuations. However, the wall-normal velocity fluctuations ($w$-fluctuations) contribute to the sweep events, in which the high-momentum fluid is directed towards the wall surface. The sweep events are known to be responsible for the turbulent skin-friction drag \citep{orlandi1994generation, choi2011turbulent}; hence, opposition control schemes attempt to suppress these events. Therefore, coherence of $w$-fluctuations with the streamwise friction velocity fluctuations too plays a key role in the efficacy of control systems employing wall-based sensing (since $w$ cannot be sensed at the wall). In order to calculate scale-dependent wall-normal coherence spectrogram ($\gamma_{{wu_\tau}}^2$), $u$ is replaced by $w$ in (\ref{eq:lcs}). The wall-normal coherence spectrograms for varying $\Delta x^+$ values between the sensor and actuator at $Re_\tau\approx$ 2000 are shown in figure \ref{fig:w_streamwise_offset_a-f}. The filled contours show the $\gamma_{{wu_\tau}}^2$ spectra for $\Delta x=0$ in all panels while the contour lines show the $\gamma_{{wu_\tau}}^2$ spectra for varying $\Delta x$; the dashed lines show the $u$-energy spectra contour level $k_x^+ \phi_{uu}^+=1.8$ and the `$+$' marker shows the peak in the energy spectra. It is evident that contrary to $\gamma_{{uu_\tau}}^2$, which demonstrates monotonic increase of coherence with wavelength and monotonic decrease of coherence with distance from the wall, $\gamma_{{wu_\tau}}^2$ spectrograms peak at $z^+ \approx 15$ and $\lambda_x^+ \approx 1700$ and coherence diminishes with increasing wavelength beyond  $\lambda_x^+ \approx 1700$. Further investigation is required to explore the implications of these features for opposition control systems. For $\Delta x=0$, the wall-normal LCS value is $\gamma_{{wu_\tau}}^2 \approx 0.4$ at the near wall energy peak ($z^+=15$ and $\lambda_x^+=810$); this coherence value gradually decreases with the increase in $\Delta x$. The relatively low wall-normal LCS at the peak energy spectra suggests again that opposition control based on wall sensor inputs can achieve only limited success, at least for fluctuating wall friction as the input. The wall-normal LCS magnitude at $z^+=15$ is shown in figures \ref{fig:w_gamma_dx_lambda} (a) as a function of $\Delta x^+$ and $\lambda_x^+$, and at $z^+=15$ and $\lambda_x^+=810$ as a function of $\Delta x^+$ in figure \ref{fig:w_gamma_dx_lambda} (b) for $Re_\tau \approx 590$ and $Re_\tau \approx 2000$. Similar to the streamwise LCS, the wall-normal LCS magnitude at the near wall peak, is constant up to $\Delta x^+ \approx 200$ for $Re_\tau \approx 590$ and up to $\Delta x^+ \approx 1000$ for  $Re_\tau \approx 2000$. The difference between the constants associated with $\gamma_{{wu_\tau}}^2$ for the two Reynolds numbers appears to be more noticeable than those associated with $\gamma_{{uu_\tau}}^2$.

\section{Conclusions}
Scale-dependent coherence between synchronized friction velocity data and streamwise and wall-normal velocity data at different wall-normal locations was explored for wall-bounded turbulent flows at $Re_\tau \approx 590$ and $2000$ with a focus on the near wall region. Coherence of the velocity signals at the peak of the energy spectrogram with the friction velocity signals is critical for exploring the feasibility of wall-based, active manipulation of near wall streaks and QSVs. The findings of this study are summarized as follows:\\[-8pt]
\begin{enumerate}[labelwidth=0.65cm,labelindent=0pt,leftmargin=0.65cm,label=(\roman*),align=left]
\item \noindent Temporal and spatial $u$-fluctuations were used to compute temporal and spatial coherence spectra. Comparison of the temporal and spatial coherence spectra revealed minimal difference between them for $z^+ \geq 10$ and $\lambda_x^+ \geq 500$; this region encompasses the inner-peak of the energy spectrogram ($z^+ \approx 15$ and $\lambda_x^+\approx 800$). Therefore, a time-resolving point sensor can adequately provide the required input signal without loss of coherence in a closed-loop drag reduction system.
\item \noindent Wall-coherent and -incoherent portions of the $u$-energy spectra were obtained and were integrated over all wavelengths to calculate wall-coherent and -incoherent portions of the streamwise turbulence intensity. It was found that the incoherent portion of the streamwise turbulence intensity is $Re$-invariant in the near-wall region (at least for the Reynolds number range $Re_\tau=590-2500$) while its coherent portion grows with $Re_\tau$ with the same rate as the total streamwise turbulence intensity. The ratio of the peak of the coherent portion of the turbulence intensity profile to the peak of the total turbulence intensity profile was therefore found to grow with $Re_\tau$. Adopting this ratio as an indicator of the wall-coherent energy to the total turbulence energy ratio, one can conclude that the effectiveness of a closed-loop drag reduction system that targets \textit{all turbulence scales} in the near wall region relying on wall-sensors can increase with $Re_\tau$. However, given that the energy associated with the near-wall streaks form an increasingly smaller component of the wall-coherent energy as Reynolds number grows, one may conclude that there will be a diminishing drag reduction from an active control scheme that only targets the near-wall streaks.
\item \noindent QSVs in the near wall region have a typical diameter in the order of $d^+=20-50$. It was shown that in order for the streamwise velocity signal and the friction velocity signal to be coherent at the inner-peak of the energy spectrogram (associated with the near wall cycle of streaks), the viscous scaled spanwise separation between the sensor and the actuator must be kept below $\Delta y^+=20$.
\item \noindent At the inner-peak of the energy spectrogram, $\gamma_{{uu_\tau}}^2 \approx 0.55$ when $\Delta x = \Delta y = 0$. Consequently, only 55\% of the $u$-energy associated with the near wall cycle turbulence is stochastically coherent with a wall-based sensor. For that reason, an active wall-based sensing and actuation control scheme for opposition control of only the near wall cycle turbulence has a limited efficacy with a theoretical upper limit of only suppressing 55\% of the turbulence at that location in the spectrogram. The LCS remains constant with increasing viscous scaled streamwise separation between the sensor and actuator ($\Delta x^+$) up to a limit, which appears to increase with Reynolds number. The LCS then rolls off for further increase in $\Delta x^+$.
\item \noindent $w$-fluctuations contribute to the sweep events, which are in turn responsible for high skin-friction drag. A wall-normal LCS of $\gamma_{{wu_\tau}}^2 \approx 0.4$ was found at the inner-peak of the $u$-energy spectrogram, which is expectedly lower than $\gamma_{{uu_\tau}}^2$ at the inner peak. This limit suggests that wall-based sensing using $u_\tau$ will have limited success in terms of estimating the $w$ fluctuations at $z^+ = 15$. Some other wall-based quantity must be sought that is better correlated with $w$ (for example pressure or in-plane gradients of $u_\tau$).
\end{enumerate}   

\section*{Acknowledgements}
The authors gratefully acknowledge the financial support of the Australian Research Council.

\bibliography{mybibfile}

\end{document}